\documentclass[final]{aipproc}
\usepackage{amsmath,subfig}
\layoutstyle{8x11single}

\begin{document}
\title{Nonuniform phases in the 't Hooft extended Nambu--Jona-Lasinio Model}

\author{J. Moreira}{
  address={Centro de F\'{i}sica Computacional, Departamento de F\'{i}sica da Universidade de Coimbra, 3004-516 Coimbra, Portugal}
}
\author{B. Hiller}{
  address={Centro de F\'{i}sica Computacional, Departamento de F\'{i}sica da Universidade de Coimbra, 3004-516 Coimbra, Portugal}
}
\author{A. A. Osipov}{
address={Centro de F\'{i}sica Computacional, Departamento de F\'{i}sica da Universidade de Coimbra, 3004-516 Coimbra, Portugal}
,altaddress={on leave from Dzhelepov Laboratory of Nuclear Problems, JINR 141980 Dubna, Russia}
}
\author{A. H. Blin}{
address={Centro de F\'{i}sica Computacional, Departamento de F\'{i}sica da Universidade de Coimbra, 3004-516 Coimbra, Portugal}
}

\begin{abstract}
The phase diagram of cold dense quark matter is studied using the 't Hooft extended Nambu-Jona--Lasinio Model applied to the light quark sector with a finite current mass for the strange quark (up and down are considered in the chiral limit). By relaxing the traditional uniformity assumptions and considering a modulated light quark condensate background we investigate the possible existence of non-uniform phases in this region of the phase diagram. The effects of changes in the coupling strengths of the model are studied and it is shown that the inclusion of flavour mixing combined with the finite current mass of the strange quark catalyses the appearance of the non-uniform phases, extending the domain for their existence.
\end{abstract}
\maketitle
\section{Introduction}

Low energy models of the Nambu--Jona-Lasinio (NJL)~\cite{Nambu:1961tp, Nambu:1961fr, Vaks:Larkin} type have long been used as a tool in the theoretical study of strongly interacting matter. The success of this approach can be tied to the fact that they share with QCD its global symmetries and incorporate a mechanism for the dynamical breaking of chiral symmetry. Explicit breaking of $U_A(1)$ can be incorporated through the inclusion of the OZI-violating 't Hooft flavour determinant \cite{'tHooft:1976fv,PhysRevD.18.2199.3, Bernard:1987gw, Bernard:1987sg, Reinhardt:1988xu}. These simple models are of particular usefulness in the low temperature/high chemical potential regime due to the sign problem which affects the more first principle approach of lattice-QCD.

As the pionic interaction with quarks (or nucleons) is attractive when the mean pion field carries a gradient it has long ago been proposed (for recent reviews see, for instance,~\cite{Broniowski:2011ef, Buballa:2014tba}) that the non-trivial balance between this effect and that of the kinetic terms could allow for the existence of non-uniform phases in cold dense strongly-interacting matter. 

Here we will outline some of the results reported in \cite{Moreira:2013ura} (to which we refer for a more in-depth discussion), pertaining to the case study of dense light-quark matter (we consider the $u$, $d$ and $s$ quarks) at zero temperature using the NJL model with a 't Hooft determinant term. A  non-vanishing current mass for the $s$ quark is taken into account whereas the chiral limit is considered for the $u$ and $d$ quarks thus validating the use of a simple chiral wave \textit{ansatz} for the light quark chiral condensates. 

\section{The model}

In the chiral limit the Euler--Lagrange equations of the model admit as an analytical solution the chiral wave \textit{ansatz} proposed in \cite{Dautry:1979bk} and given by the following quark orbitals and energy levels:
\begin{align}
\label{ansatz}
\langle\overline{\psi_l}\psi_l\rangle&=\frac{h_l}{2} \mathrm{cos}(\vec{q}\cdot\vec{r}), \quad
\langle\overline{\psi_l} i \gamma_5\tau_3\psi_l\rangle=\frac{h_l}{2} \mathrm{sin}(\vec{q}\cdot\vec{r}),\nonumber\\ 
E^{\pm}&=\sqrt{M^2+p^2+\frac{q^2}{4}\pm\sqrt{\left(\vec{p}\cdot\vec{q}\right)^2+M^2 q^2}},
\end{align}
Here $M$ and $\boldsymbol{p}$ denote the dynamical mass and the momentum of the quark, $\boldsymbol{q}$ is the wave vector of the condensate modulation and  $\tau_3$ is the Pauli matrix acting in isospin space. Note that the $E^-$ branch has a lower energy than $E^+$ and thus this branch is preferably occupied. For the $s$ quark no modulation is considered (uniform strange condensate background).

In the mean field approximation the application of techniques of Ref.~\cite{Osipov:2003xu} yields the thermodynamic potential of the model as:
\begin{align}
\label{Omega}
\Omega=&V_{st}+\frac{N_c}{8\pi^2}\left(J_{-1}(M_u,\mu_u,q)+J_{-1}(M_d,\mu_d,q)+J_{-1}(M_s,\mu_s,0)\right)\nonumber\\
V_{st}=&\frac{1}{16}\left.\left(4G\left(h_u^2+h_d^2+h_s^2\right)+\kappa h_u h_d h_s\right)\right|^{M_i}_0,
\end{align}
where $h_i$ ($i=u,d,s$) are twice the quark condensates. The coupling strengths $G$ and $\kappa$ refer respectively to the NJL and the 't Hooft determinant terms. The bosonization of the model through the introduction of auxiliary variables enables the evaluation of the fermionic part which is rendered quadratic. The stationary phase contribution coming from the integration over the auxiliary fields is given by $V_{st}$ while the contribution coming from the fermionic path integral is given by the $J_{-1}$ integrals. The notation $|^M_0$ refers to the subtraction of the quantities evaluated at the dynamical masses by their value evaluated at $M=0$ \cite{Hiller:2008nu}.
Using a regularization kernel corresponding to two Pauli-Villars subtractions in the integrand \cite{Osipov:1985}, previously used for instance in \cite{Osipov:2006xa, Osipov:2007mk}, namely $\rho\left(s\Lambda^2\right)=1-(1+s\Lambda^2)\mathrm{exp}(-s\Lambda^2)$, the vacuum and medium contributions, $J^{\mathrm{Vac}}_{-1}$ and $J^{\mathrm{Med}}_{-1}$, can be written explicitly as:
\begin{align}
J_{-1}=&J^{Vac}_{-1}+J^{Med}_{-1},\nonumber\\
J^{Vac}_{-1}=&\int\frac{\mathrm{d}^4 p_E}{(2\pi)^4}\int^\infty_0 \frac{\mathrm{d}s}{s}\rho\left(s\Lambda^2\right)8\pi^2e^{-s\left(p_ {0\,E}^2+p_\perp^2\right)}
\left.\left(e^{-s\left(\frac{q}{2}+\sqrt{M^2+p_z^2}\right)^2}+e^{-s\left(\frac{q}{2}-\sqrt{M^2+p_z^2}\right)^2}\right)\right|^{M,q}_{0,0},\nonumber\\
J^{Med}_{-1} =&-\int\frac{\mathrm{d}^3p}{(2\pi)^3}8\pi^2T\left.\left(\mathcal{Z}^+_++\mathcal{Z}^+_-+\mathcal{Z}^-_++\mathcal{Z}^-_-\right)\right|^{M,q}_{0,0}+C(T,\mu),\nonumber\\
\mathcal{Z}^\pm_\pm =&\mathrm{log}\left(1+e^{-\frac{E^\pm\mp\mu}{T}}\right)-\mathrm{log}\left(1+e^{-\frac{E_\Lambda^\pm\mp\mu}{T}}\right)-
\frac{\Lambda^2}{2T E_\Lambda^\pm}\frac{e^{-\frac{E_\Lambda^\pm\mp\mu}{T}}}{1+e^{-\frac{E_\Lambda^\pm\mp\mu}{T}}},\nonumber\\
C(T,\mu)=&\int\frac{\mathrm{d}^3p}{(2\pi)^3}16\pi^2T
\, 
\mathrm{log}\left(\left(1+e^{-\frac{|\boldsymbol{p}|-\mu}{T}}\right)\left(1+e^ { 
-\frac{|\boldsymbol{p}|+\mu}{T}}\right)\right)
\end{align}
where $E_\Lambda^{\pm}=\sqrt{\left(E^{\pm}\right)^2+\Lambda^2}$. The $z$-axis was chosen to coincide with the direction of modulation wave vector. The $|^{M,q}_{0,0}$ notation refers to the subtraction of the same quantity evaluated for a uniform gas of massless quarks ($M=0$ and $q=0$).
The superscript and subscript $\pm$ in the definition of $\mathcal{Z}$ refer respectively to the energy branch and the sign in front of the chemical potential in the exponent. The $C(T,\mu)$ term is added for thermodynamic consistency \cite{Hiller:2008nu} ensuring that the correct asymptotic behaviour is obtained in the $T\rightarrow \infty$ limit.

The minimization of the thermodynamical potential with respect to $M$ and $q$ has to be done self-consistently with the resolution of the stationary phase equations:
\begin{align}
\label{StaEq}
\left\{
\begin{array}{l}
m_u-M_u=G h_u +\frac{\kappa}{16}h_d h_s\\
m_d-M_d=G h_d +\frac{\kappa}{16}h_u h_s\\
m_s-M_s=G h_s +\frac{\kappa}{16}h_u h_d
\end{array}	
\right.,
\end{align}
where $m_i$ stands for the current masses of the quarks.

\section{Results}
In our study the current masses were set to $m_u=m_d=0$, $m_s=186~\mathrm{MeV}$. The remaining three parameters of the model ($G$, $\kappa$ and $\Lambda$) can be reduced to two, the 't Hooft coupling strength, $\kappa$, and the dimensionless curvature, $\tau=N_c G \Lambda^2/(2\pi^2)$, by fitting $\Lambda$ to reproduce a reasonable value for the vacuum dynamical mass of the light quarks ($M_l=300~\mathrm{MeV}$).

With no OZI-violating term, the light and strange sectors are decoupled and in the chiral limit $\tau=1$ is the critical value above which dynamical chiral symmetry breaking occurs:  as a crossover for $1<\tau<1.23$ and as a first order transition for higher values. The transitions are independent for the  light and strange sector, with the latter shifted to higher $\mu$ due to the current $s$-quark mass (both occur at values of $\mu$ close to the vacuum dynamical mass $M^{\rm vac}_i$).

For high enough chemical potential an energetically favourable modulated solution always emerges. This solution goes asymptotically to $\lim_{\mu\rightarrow\infty}\left\{h,q\right\}=\{0,2\mu\}$ and becomes degenerate with the trivial one.

Above $1.23<\tau<1.53$, a window for energetically favourable finite-$q$ solutions appears in the vicinity of the first-order transition associated with the condensate of the light quarks and the usual transition to a vanishing condensate is superseded by a transition to a finite but modulated condensate: $\left\{h_l,0\right\}\rightarrow\left\{h_l^\prime,q\right\}$. The window of existence of this phase ends with the light condensate going continuously to zero (note that if the condensate vanishes its modulation is ill defined). The extent of this windows is increased for higher values of $\tau$ and for $\tau>1.53$ it merges with the one at higher chemical potentials.
\begin{figure}[htb]
\centerline{\includegraphics[width=1\textwidth]{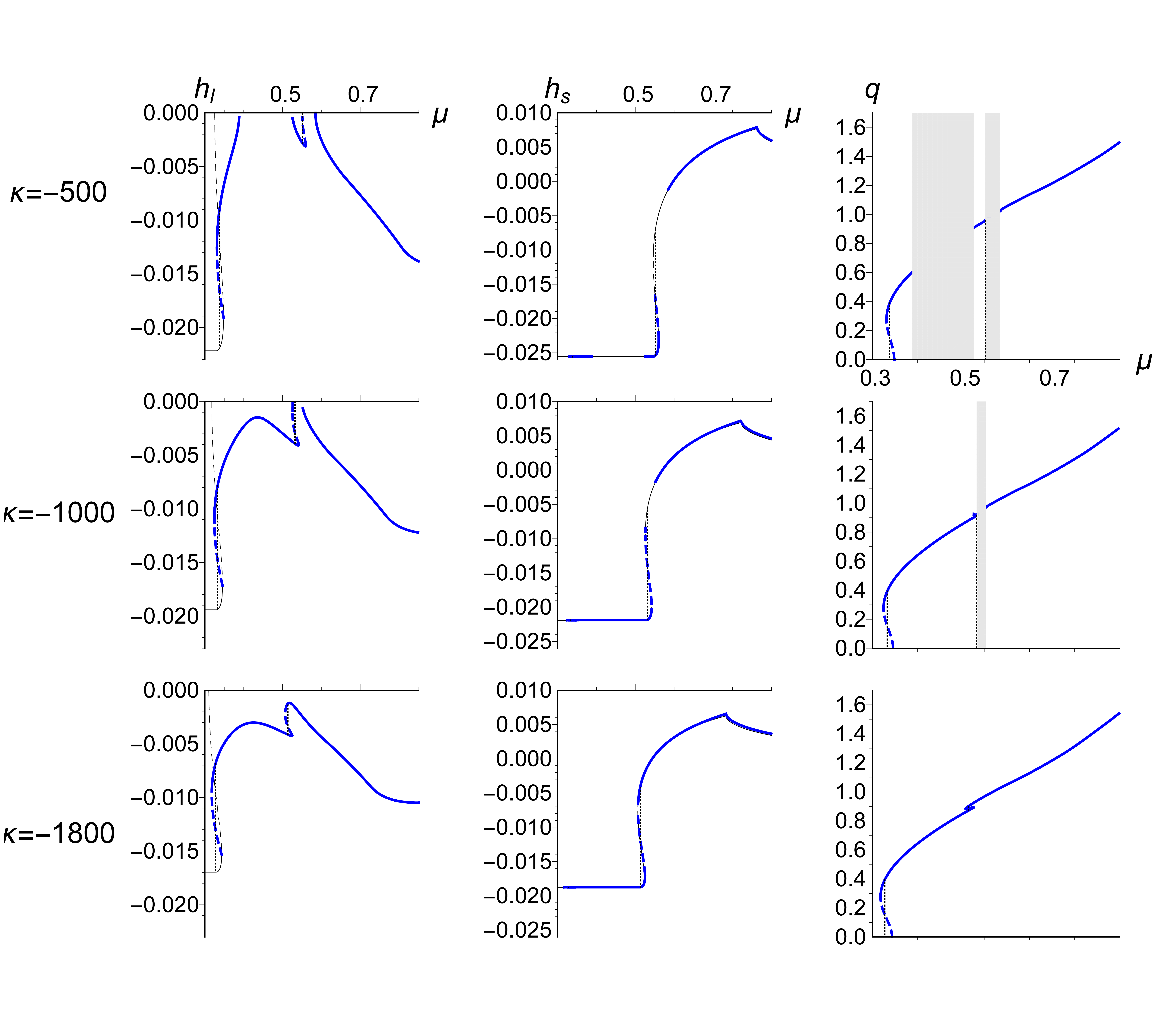}}
\caption{From left to right: the chemical potential dependence of the zero temperature solutions for the light condensate, $h_l$, the strange condensate, $h_s$, and for the modulation wave vector $q$. From top to bottom: each row corresponds to a different choice of the 't Hooft coupling strength ($\left[\kappa\right]=\mathrm{GeV}^5$).
Thicker lines denote the inhomogeneous solutions. In the grey regions the value of $q$ is undetermined, since $h_l=0$.}
\label{grafpainelNJLHSU3}
\end{figure}

The inclusion of the 't Hooft determinant term couples the three flavour sectors leading to several different scenarios, at a fixed curvature, depending on the the value of its coupling strength. Some examples of this can be seen in Fig.~\ref{grafpainelNJLHSU3} for the $\tau=1.4$ case. 
For $-\kappa>290~\mathrm{GeV}^{-5}$ a new window of energetically favourable modulated solutions appears in the vicinity of $\mu\sim M^{\rm vac}_s$  (see the shark-fin shape for $h_l$  in the first row of Fig.~\ref{grafpainelNJLHSU3}). There are therefore three intervals of chemical potential where the global minimum of the thermodynamical potential corresponds to a modulated solution. The boundaries of these intervals corresopnd to two first order transitions and three-crossovers, with the latter  corresponding to the (dis)appearance of a non-vanishing $h_l$. The two first order transitions exclude the occurrence of the $q=0$ transitions as they occur at slightly lower/higher chemical potential for the one taking place near $M_l^{\rm Vac}$/$M_s^{\rm vac}$. A zoom of the behaviour of the chiral condensates near the transitions can be seen in Fig.~\ref{SolhlsNJLHtau14K500zoom} for the $-\kappa=500~\mathrm{GeV}^{-5}$ case. 

By increasing the coupling of the 't Hooft interaction even further we can expand the extent of these chemical potential windows with finite-$q$ solutions and for $-\kappa>935~\mathrm{GeV}^{-5}$ the first two merge.  As a consequence two crossover transitions disappear (see second row in Fig.~\ref{grafpainelNJLHSU3}). For $-\kappa>1660~\mathrm{GeV}^{-5}$ the last transition is no longer to homogeneous solution as there is an overlap of two distinct modulated solutions and the first-order transition occurs between two phases with finite $q$ (third row of Fig.~\ref{grafpainelNJLHSU3} and Fig. \ref{grafSolNJLHtau14K1000}).

\begin{ltxfigure}[htb]
\subfloat[\label{grafSolhlNJLHtau14K500zoomI}]{\includegraphics[width=0.25\textwidth]{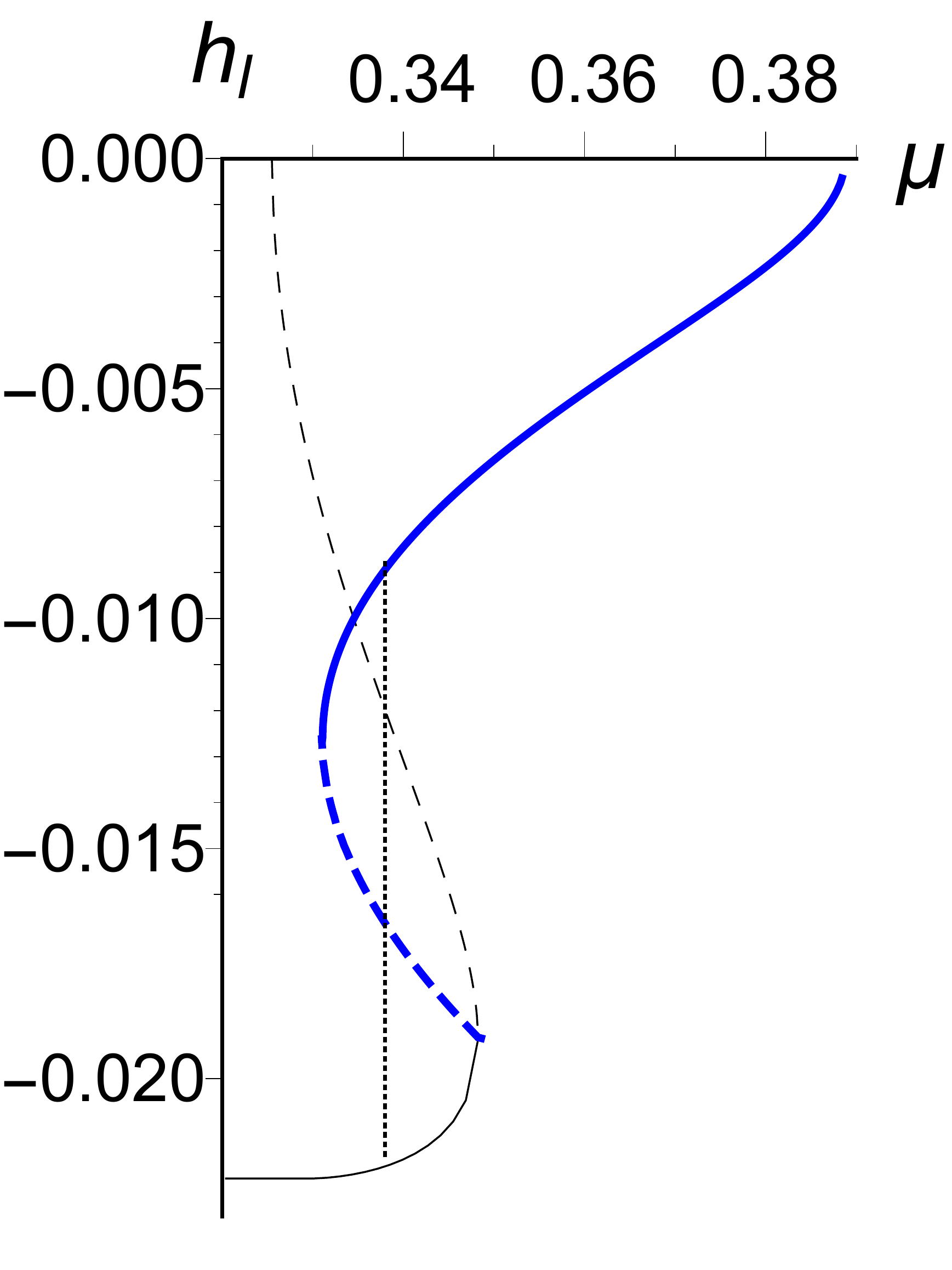}}
\subfloat[\label{grafSolhsNJLHtau14K500zoomI}]{\includegraphics[width=0.25\textwidth]{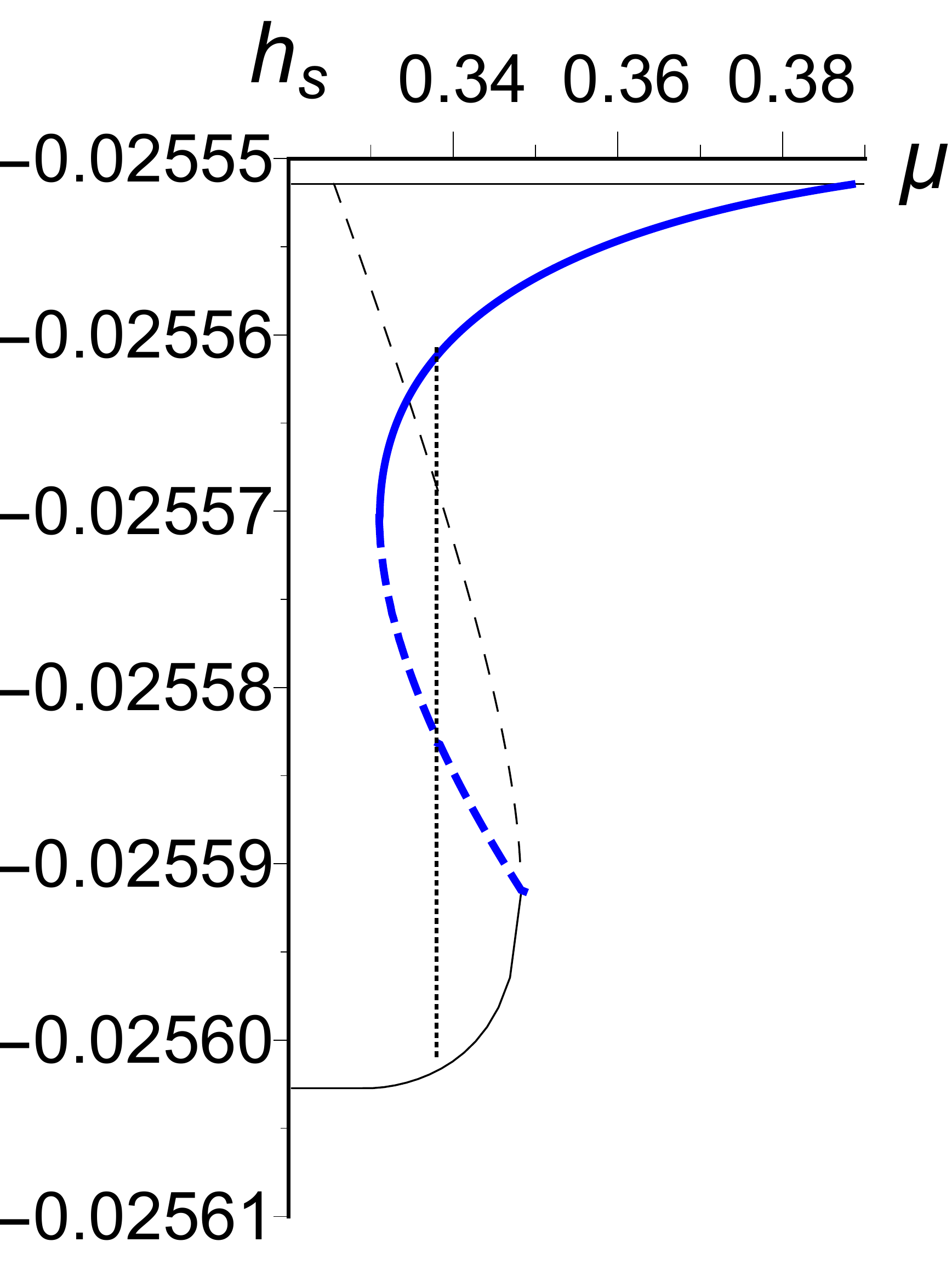}}
\subfloat[\label{grafSolhlNJLHtau14K500zoomII}]{\includegraphics[width=0.25\textwidth]{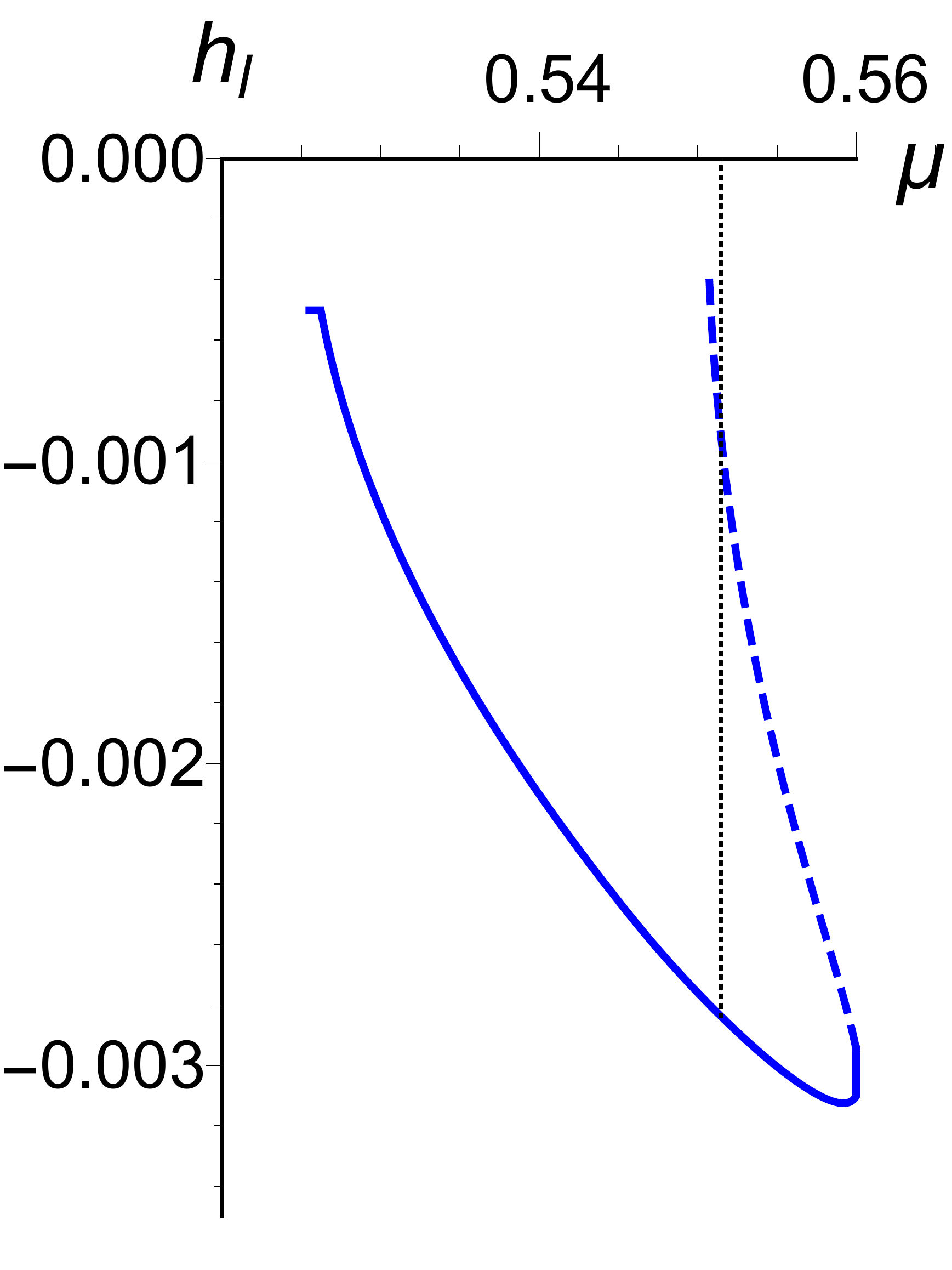}}
\subfloat[\label{grafSolhsNJLHtau14K500zoomII}]{\includegraphics[width=0.25\textwidth]{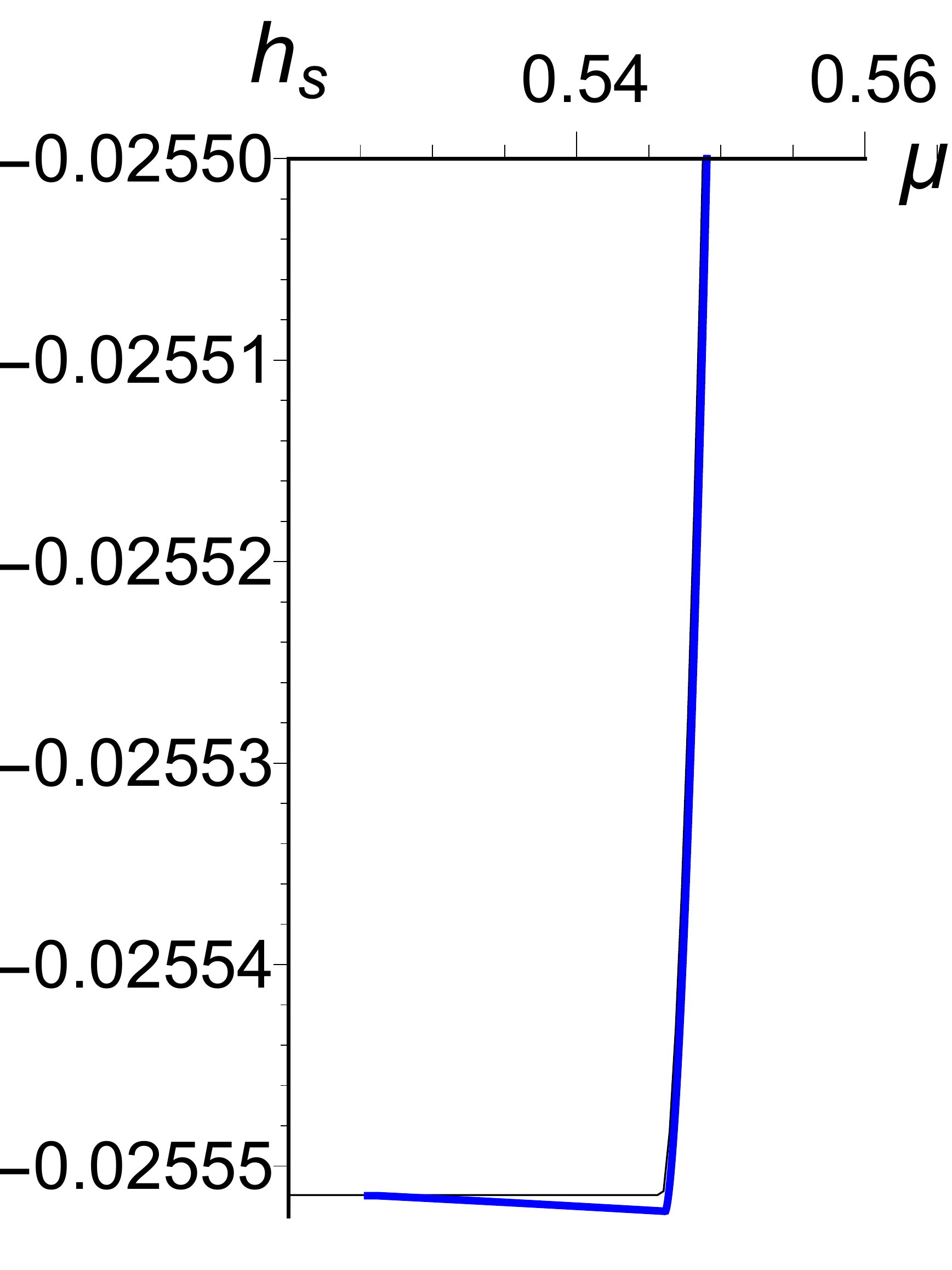}}
\caption{Zoom for the chemical potential dependence of the chiral condensate solutions, with thicker lines referring to the finite-$q$ solutions, in the vicinity of the transitions (marked by the vertical dotted lines), case of $\kappa=-500~\mathrm{GeV}^{-5}$.}
\label{SolhlsNJLHtau14K500zoom}
\end{ltxfigure}

The main results can be summarized in the form of critical chemical potentials associated with the (dis)appearance of the inhomogeneous phases shown in Fig.~\ref{PotQuimCrit}.

\begin{ltxfigure}[htb]
\center
\subfloat[\label{grafSolqNJLHtau14K500zoom}]{\includegraphics[width=0.25\textwidth]{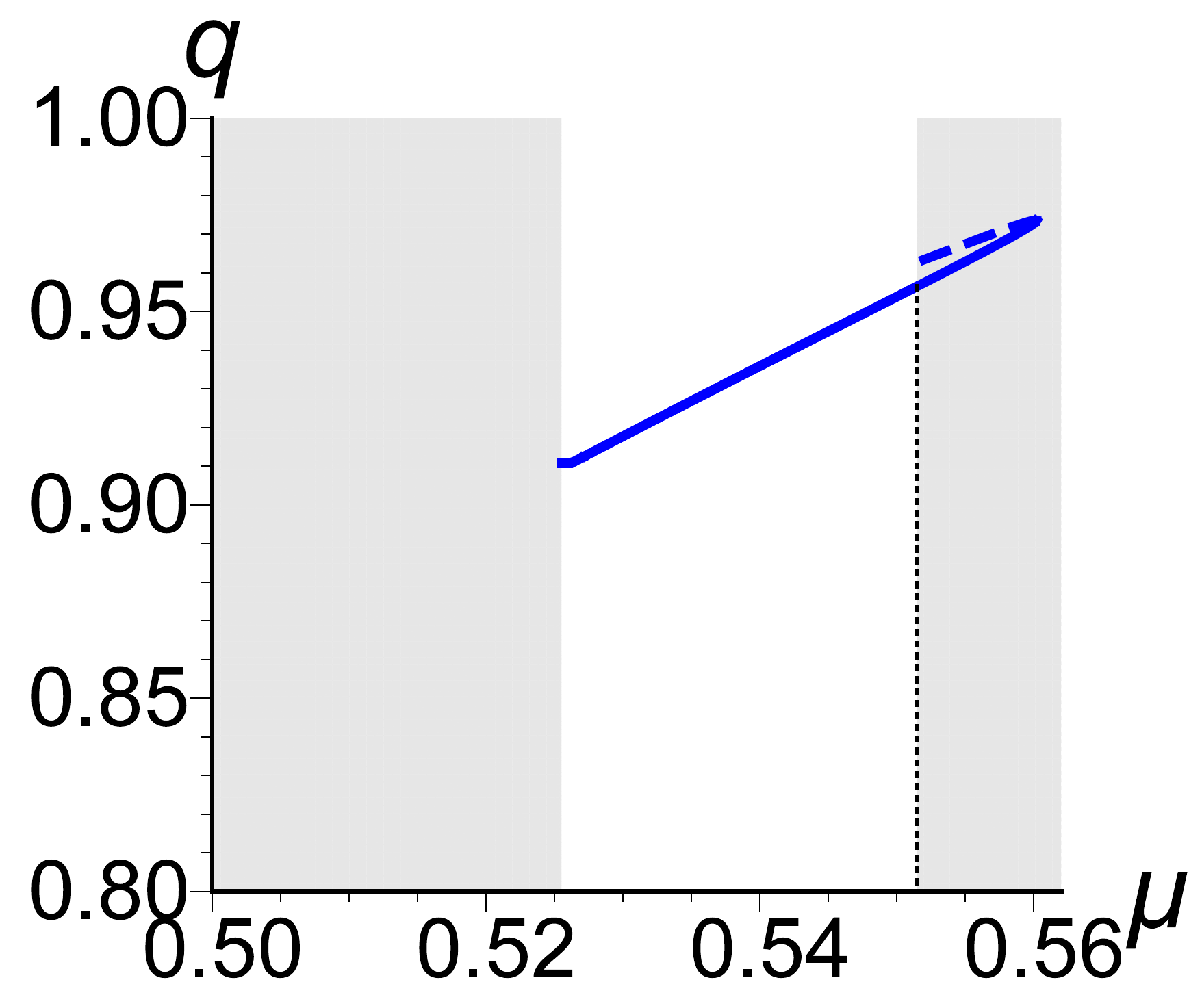}}
\subfloat[\label{grafSolqNJLHtau14K1000zoom}]{\includegraphics[width=0.25\textwidth]{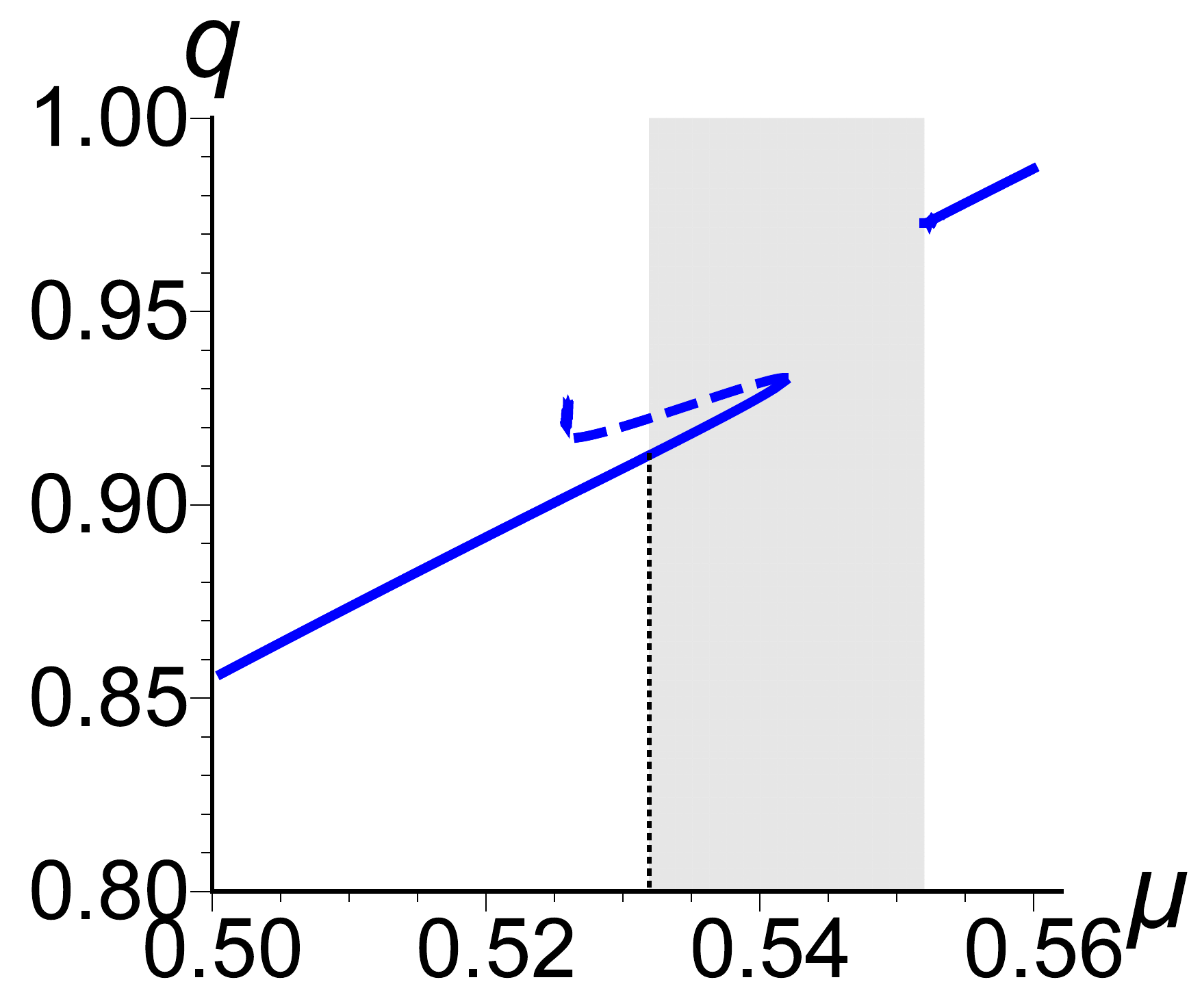}}
\subfloat[\label{grafSolqNJLHtau14K1800zoom}]{\includegraphics[width=0.25\textwidth]{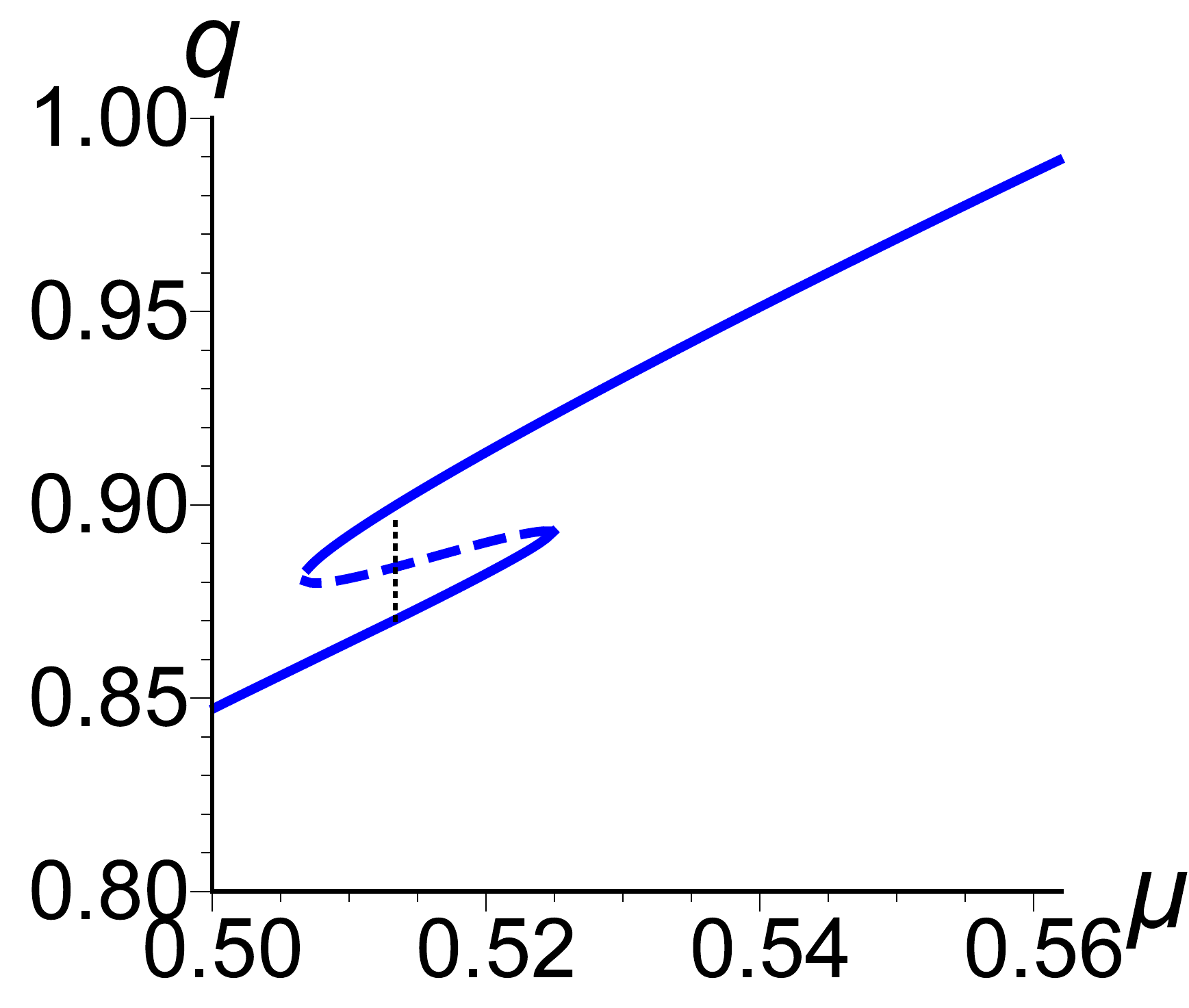}}
\caption{Zoom of the solutions for $q$ in the chemical potential window close to $M^{\mathrm{Vac}}_s$ for $\kappa=-500,-1000$, and $-1800~\mathrm{GeV}^5$ (from left to right)
, showing the merging of the solution branches for strong enough flavor mixing.
}
\label{grafSolNJLHtau14K1000}
\end{ltxfigure}

\begin{ltxfigure}[htb]
\center
\subfloat[\label{grafPotQuimCritVsTauNJLCW}]{\includegraphics[width=0.3\textwidth]{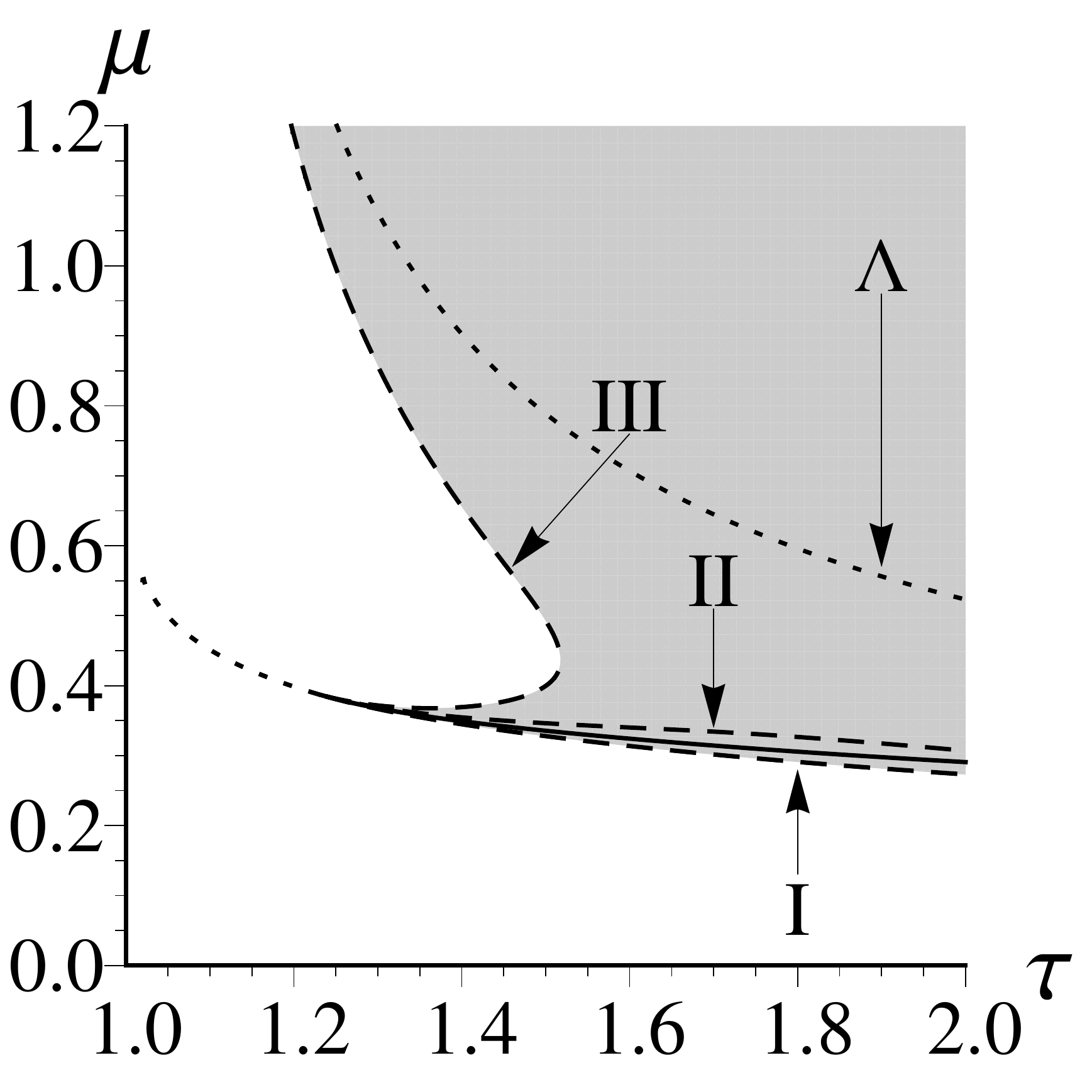}}
\subfloat[\label{grafPotQuimCritVsKNJLHCWtau14}]{\includegraphics[width=0.3\textwidth]{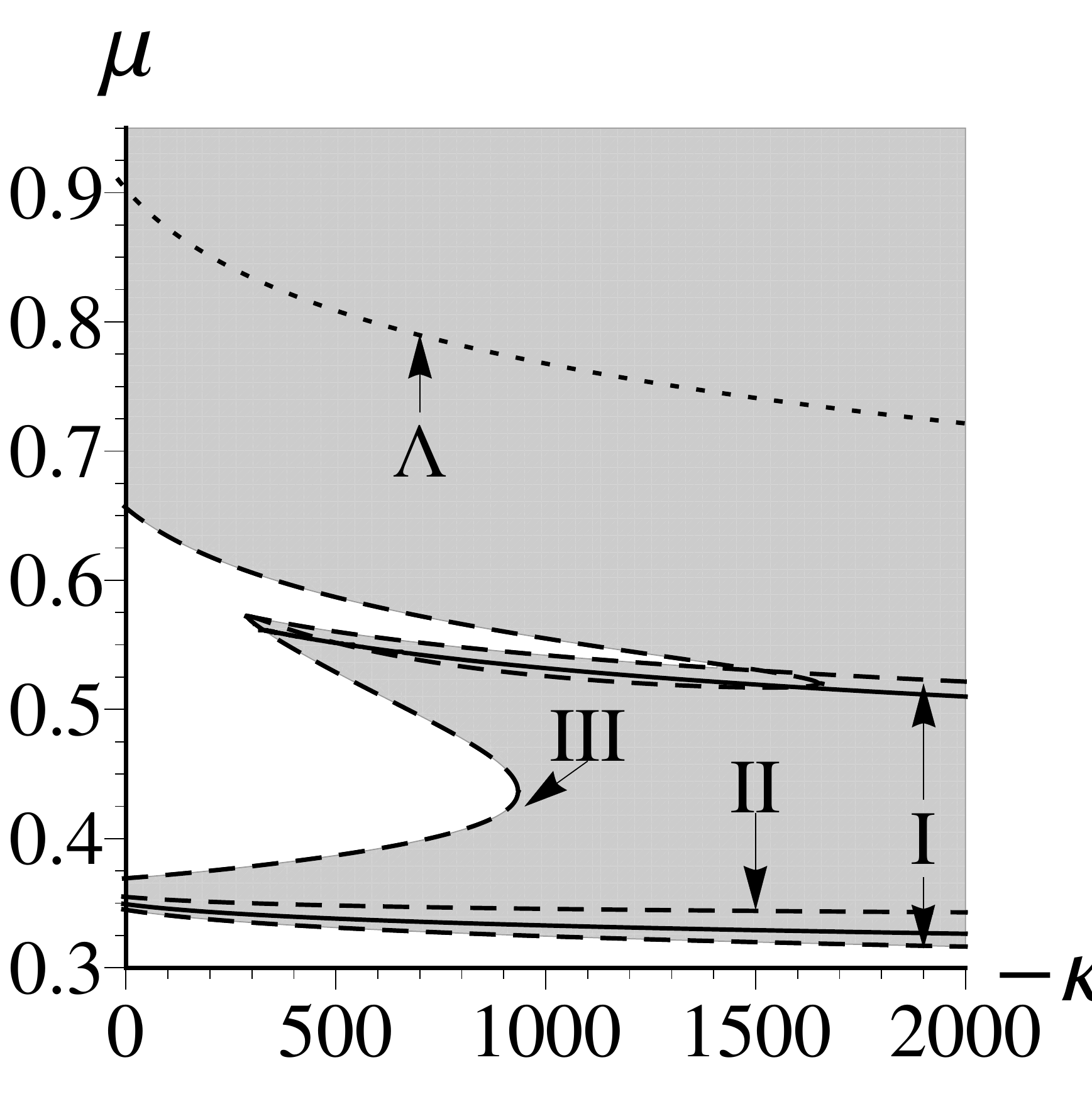}}
\subfloat[\label{grafAuxPotQuimSolNJLHCW}]{\includegraphics[width=0.3\textwidth ]{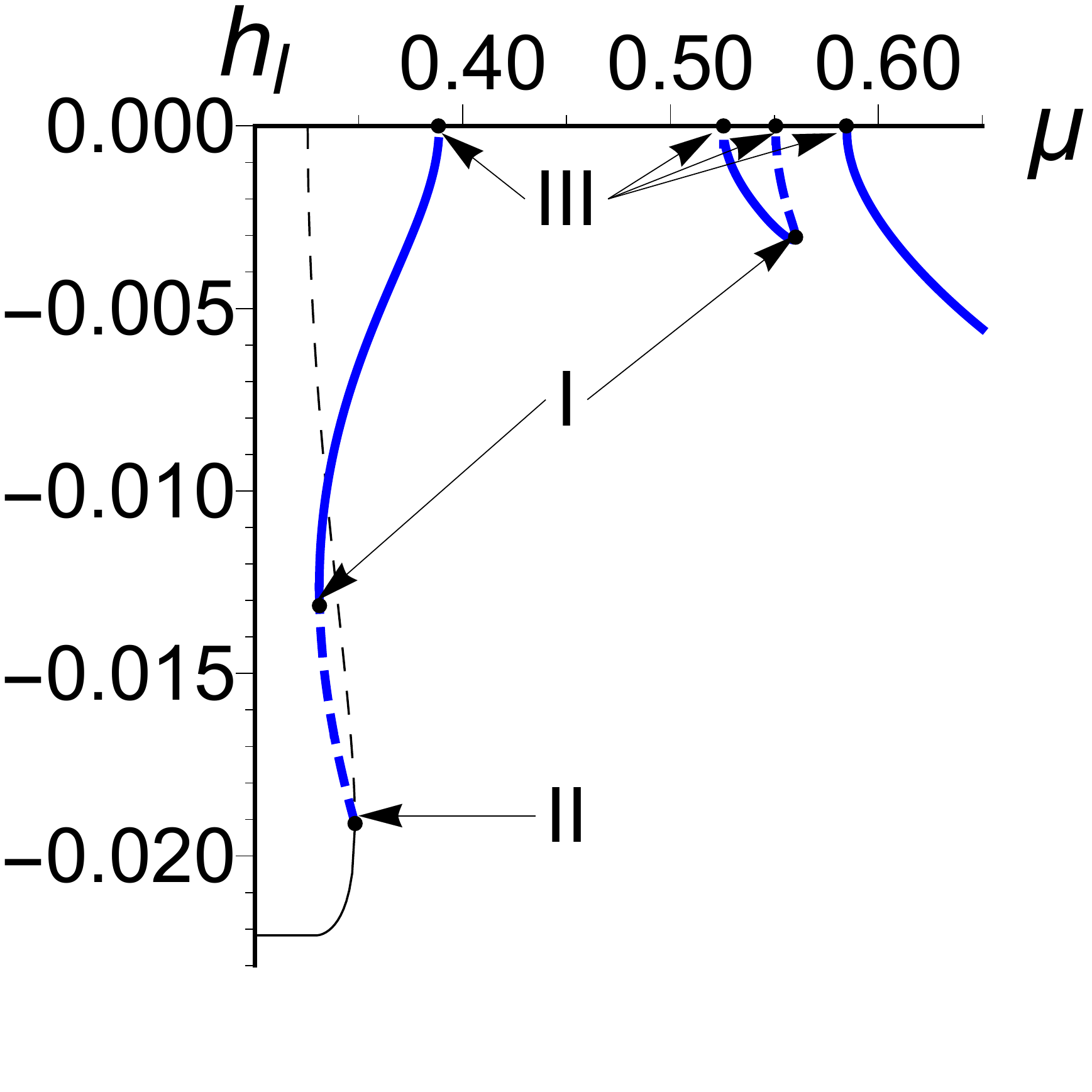}}
\caption{In \ref{grafPotQuimCritVsTauNJLCW} we present the critical chemical potentials as a function of $\tau$ in the NJL model and in \ref{grafPotQuimCritVsKNJLHCWtau14} as functions of $\kappa$ ($[\kappa]=\mathrm{GeV}^5$) in the NJLH model for the $\tau=1.4$ case. The upper dotted line corresponds to the cut-off $\Lambda$. The chemical potential of the first order transitions are marked with the full black lines. Dashed lines indicate the borders of the region where the finite-$q$ solutions exist. We distinguish between three types of critical chemical potentials and an example for this distinction in the $\kappa=-500~\mathrm{GeV}^{-5}$ case appears in Fig.~\ref{grafAuxPotQuimSolNJLHCW} (type I with finite $q$ and $h_l$, type II with finite $h_l$ and vanishing $q$, and type III with finite $q$ but vanishing $h_l$).}
\label{PotQuimCrit}
\end{ltxfigure}

We conclude that the flavour mixing effect of the 't Hooft determinant combined with the inclusion of a finite current mass for the strange quark acts as a catalyst for the appearance of globally stable inhomogeneous solutions in cold quark matter by coupling the behaviour of the light and strange sectors. Further extensions of this model to include, for instance, the eight quark interactions term \cite{Osipov:2005tq, Osipov:2006ns} are straightforward and can provide an even richer picture for the phase diagram of cold strongly interacting matter. 
	
\begin{theacknowledgments}
This work has been supported by the Centro de F\'{i}sica Computacional of the University of Coimbra, Funda\c{c}\~ao para a Ci\^encia e Tecnologia, project: CERN/FP/116334/2010, developed under the iniciative QREN, financed by UE/FEDER through COMPETE - Programa Operacional Factores de Competitividade and the grant SFRH/BPD/63070/2009/. This research is part of the EU Research Infrastructure Integrating Activity Study of Strongly Interacting Matter (HadronPhysics3) under the 7th Framework Programme of EU, Grant Agreement No. 
283286.
\end{theacknowledgments}
\bibliographystyle{aipproc}
\bibliography{JMoreiraQCHS2014v2}
\end{document}